\documentclass[aps,twocolumn,
superscriptaddress,
footinbib,
prl
]{revtex4-2}
\usepackage{amsmath,amssymb,bm}
\usepackage{graphicx}
\graphicspath{{./}{img/}}
\usepackage{epstopdf}
\usepackage{latexsym}
\usepackage[normalem]{ulem} 
\usepackage[caption=false]{subfig}
\usepackage[usenames,dvipsnames]{color}
\usepackage{hyperref}
\usepackage{natbib}
\usepackage{bbold}

\usepackage{xcolor}
\hypersetup{
	colorlinks,
	linkcolor={blue!40!black},
	citecolor={blue!40!black},
	urlcolor={blue!40!black}
}

\usepackage{amsthm}
\newtheorem{theorem}{Theorem}
\newtheorem*{theorem*}{Theorem}

\usepackage[utf8]{inputenc} 

\usepackage{newunicodechar}
\newunicodechar{ρ}{\rho}
\newunicodechar{σ}{\sigma}
\newunicodechar{λ}{\lambda}
\newunicodechar{Λ}{\Lambda}
\newunicodechar{μ}{\mu}
\newunicodechar{ν}{\nu}
\newunicodechar{ψ}{\psi}
\newunicodechar{ϕ}{\phi}
\newunicodechar{φ}{\varphi}
\newunicodechar{π}{\pi}

\usepackage{xparse, xspace, mathtools, braket}
\usepackage{csquotes}

\usepackage{physics}

\renewcommand{\tr}{\Tr}
\DeclarePairedDelimiter\autobracket{(}{)}
\newcommand{\br}[1]{\autobracket*{#1}}
\newcommand{\HS}{\mathcal{H}}
\NewDocumentCommand{\BO}{o}{\mathcal{B}\br{\IfValueTF{#1}{#1}{\HS}}}
\NewDocumentCommand{\DM}{o}{\mathcal{D}\br{\IfValueTF{#1}{#1}{\HS}}}
\DeclarePairedDelimiterX{\infdivx}[2]{(}{)}{
	#1\;\delimsize|\delimsize|\;#2
}

\NewDocumentCommand{\ent}{d()g}{\ensuremath{H\br{\IfValueTF{#1}{#1}{#2}}}}
\NewDocumentCommand{\qent}{d()g}{\ensuremath{H\br{\IfValueTF{#1}{#1}{#2}}}}

\NewDocumentCommand{\es}{m o}{
	\IfNoValueTF{#2}{
		\mathbb{\uppercase{#1}}^{\lowercase{#1}}
	}{
		\mathbb{\uppercase{#1}}^{#2}
	}
}
\newcommand{\esbraket}[2]{\bra{\es{#1}}\ket{\es{#2}}}

\newcommand{\xeket}{\ket{\es{x}}}
\newcommand{\xpeket}{\ket*{\es{X'}}}

\newcommand{\zeket}{\ket{\es{z}}}

\newcommand{\zpeket}{\ket*{\es{Z'}}}

\NewDocumentCommand{\coverlapb}{m}{\ensuremath{c_{#1}}}
\NewDocumentCommand{\coverlap}{m m}{\coverlapb{#1 #2}}

\begin{document}

\title{Experimentally accessible bounds on distillable entanglement from entropic uncertainty relations}

\date{\today}

\author{Bjarne Bergh}
\affiliation{Kirchhoff-Institut f\"{u}r Physik, Universit\"{a}t Heidelberg, Im Neuenheimer Feld 227, 69120 Heidelberg, Germany}
\author{Martin G\"{a}rttner}
\affiliation{Kirchhoff-Institut f\"{u}r Physik, Universit\"{a}t Heidelberg, Im Neuenheimer Feld 227, 69120 Heidelberg, Germany}
\affiliation{Physikalisches Institut, Universit\"at Heidelberg, Im Neuenheimer Feld 226, 69120 Heidelberg, Germany}
\affiliation{Institut f\"ur Theoretische Physik, Ruprecht-Karls-Universit\"at Heidelberg, Philosophenweg 16, 69120 Heidelberg, Germany}

\begin{abstract}
Entanglement is not only the resource that fuels many quantum technologies but also plays a key role for some of the most profound open questions of fundamental physics. Experiments controlling quantum systems at the single quantum level may shed light on these puzzles. However, measuring, or even bounding, entanglement experimentally has proven to be an outstanding challenge, especially when the prepared quantum states are mixed.
We use entropic uncertainty relations for bipartite systems to derive measurable lower bounds on distillable entanglement. We showcase these bounds by applying them to physical models realizable in cold-atom experiments. 
The derived entanglement bounds rely on measurements in only two different bases and are generically applicable to any quantum simulation platform.
\end{abstract}

\maketitle

\textit{Introduction.}--
Quantum entanglement is known to lie at the heart of many physical phenomena \cite{Amico2008}.
For example, in quantum statistical mechanics \cite{Deutsch1991, Rigol2008} and quantum field theory \cite{Berges2018} entanglement explains how time-evolving quantum systems become locally thermal. In condensed matter theory, entanglement allows us to characterize topological states of matter \cite{Kitaev2006}, and it may explain the information paradox in quantum gravity \cite{Harlow2016}.
At the same time entanglement is also the reason why many of these phenomena remain poorly understood as it makes simulating strongly correlated quantum systems numerically prohibitively hard.

Rapid progress on experimental techniques over the past decades has led to a range of readily available platforms that can emulate strongly interacting quantum systems in highly controlled settings, beyond what is possible on classical computers \cite{Georgescu2014}.
If we want to harness such quantum simulation experiments for addressing the questions mentioned above, methods for detecting entanglement in experimental data are needed \cite{Guehne2009}. This poses a challenge as the entanglement entropy, the prime entanglement measure for pure states, cannot be extracted from a few local observables. Its determination generally requires the full reconstruction of the prepared quantum state, which is only feasible for very small system sizes.
Recently, two ways for bypassing this problem have been proposed and implemented successfully. By preparing multiple identical copies of a many-body system and interfering them, R\'enyi entanglement entropies become accessible \cite{Ekert2002, Jaksch2004, Daley2012, islam_measuring_2015, kaufman_quantum_2016}. The second approach to entanglement quantification is based on random measurements \cite{elben_renyi_2018, brydges_probing_2019}. These schemes rely on the possibility to prepare almost pure states, which is \textit{a priori} not given in experiments due to noise and decoherence. 
In the case of general mixed states entanglement entropy is not a valid entanglement quantifier any more. Various ways exist to generalize entanglement entropy to mixed states \cite{Plenio2007}, one of them being distillable entanglement defined as the maximum number of Bell pairs that one can on average obtain when repeatedly preparing a state. Such quantities are challenging to evaluate even numerically for more than a few qubits \cite{Plenio2007}. No methods for directly measuring mixed state entanglement are known. What is experimentally accessible are lower bounds on entanglement measures \cite{Guehne2009, Bavaresco2018, schneeloch_quantifying_2018, schneeloch_quantifying_2019, dai_experimentally_2020}. However, these often lack tightness and scalablility to large system sizes.

Here we derive a measurable lower bound on coherent information, which in turn is a lower bound on distillable entanglement \cite{devetak_distillation_2005}. Our bound is based on entropic uncertainty relations for bipartite systems \cite{coles_entropic_2017} and requires measuring in only two different local bases.
The basic principle behind these relations goes back to the EPR paradox \cite{Einstein1935}. In a bipartite system, measurements made on one subsystem allow the prediction of the outcomes of measurements on the other one. If such prediction is possible, beyond a certain degree of precision, for two measurements that have no common eigenstates, then the state must be entangled. Entropic uncertainty relations turn this into a quantitative statement using conditional entropies and thus into a bound on coherent information \cite{coles_entropic_2017}. We derive an entropic entanglement bound that improves upon previously known bounds and thereby enables entanglement detection in typical cold-atom experiments.

\textit{Entanglement bounds.}--
The quantum state of a bipartite system is described by the density operator $\hat{\rho}_{AB}$ on a Hilbert space $\mathcal{H}_A\otimes\mathcal{H}_B$ with local Hilbert spaces $\mathcal{H}_A$ and $\mathcal{H}_B$, assumed to be finite-dimensional here. We consider local non-degenerate projective measurements.
% , characterized by basis states $\xeket$ for outcomes $x$.
The joint probability for obtaining outcome $x_A$ upon measuring in basis $X=\{\xeket\}$ in subsystem $A$ and $x_B$  upon measuring in basis $X'=\{\xpeket\}$ in $B$ is $P_{XX'}(x_A,x'_B)$. We quantify the average uncertainty about $x_A$ given the outcome $x'_B$ by the classical conditional entropy $\qent(X_A|X'_B)=\sum_{x'_B}P_{X'}(x'_B)H(X_A|X'_B\!=\!x'_B)$, where $P_{X'}(x'_B)$ is the marginal distribution over outcomes in $B$ and $H(X_A|X'_B\!=\!x'_B)=-\sum_{x_A}P_{XX'}(x_A|x'_B)\log\left[P_{XX'}(x_A|x'_B)\right]$ is the Shannon entropy of the conditional distribution $P_{XX'}(x_A|x'_B) = P_{XX'}(x_A,x'_B)/P_{X'}(x'_B)$. If the two subsystems are strongly correlated such that precise inference of outcomes is possible, the conditional entropy becomes small.

If we additionally measure in a second pair of bases $Z=\{\zeket\}$ in $A$ and $Z'=\{\zpeket\}$ in $B$, the following entropic uncertainty equation holds \cite{berta_uncertainty_2010, coles_entropic_2017}:
\begin{equation}
\label{eq:urelMU}
    H(X_A|X'_B) + H(Z_A|Z'_B) \geq  q_{\rm MU} + H(A|B)
\end{equation}
where $q_{\rm MU} = - \log(\max_{x, z} \abs{\esbraket{z}{x}}^2 )$ is the complementarity factor, introduced by Maassen and Uffink \cite{maassen_generalized_1988}, which quantifies how incompatible two measurements are. $q_{\rm MU}$ only depends on the measurement bases chosen in subsystem $A$. The quantity $H(A|B)=H(\hat{\rho}_{AB})-H(\hat{\rho}_B)$ is the \emph{quantum} conditional entropy, where $\qent{\hat{\rho}} = - \tr[\hat{\rho} \log(\hat{\rho})]$ is the von Neumann entropy and $\hat{\rho}_A=\tr_B(\hat{\rho}_{AB})$ the reduced state of subsystem $A$. $I_{B\rangle A}=-H(A|B)$ is called the coherent information. It is a lower bound on distillable entanglement \cite{devetak_distillation_2005} and for pure states $\hat{\rho}_{AB}=\ket{\psi}\bra{\psi}$ it  reduces to the entanglement entropy. Thus, Eq.~\eqref{eq:urelMU} gives a lower bound on distillable entanglement involving classical conditional entropies of measurable distributions and the complementarity factor, which can be calculated. 
Intuitively, if the measurement outcomes are strongly correlated between $A$ and $B$ (i.e.\  $H(X_A|X'_B)$ and $H(Z_A|Z'_B)$ are small) and $X$ and $Z$ are strongly incompatible (i.e.\ $q_{\rm MU}$ is large), then one detects a large amount of entanglement.

The example of a maximally entangled state measured in mutually unbiased bases (MUBs) illustrates that the inequality Eq.~\eqref{eq:urelMU} can indeed give tight entanglement bounds. Consider a system of local Hilbert space dimension $d$ in the pure state $\ket{\psi} = \sum_x \sqrt{1/d} \xeket_A \otimes \xeket_B$, for which $-H(A|B)=\log(d)$. Upon measuring both subsystems in the basis $\xeket$, one obtains the distribution $P_{XX}(x_A,x_B)=\delta_{x_A,x_B}/d$, and thus perfectly correlated outcomes leading to $H(X_A|X_B)=0$. Strikingly, measuring in the Fourier transformed basis $\zeket=\sum_x \sqrt{1/d} \exp(2 \pi i xz/d)\xeket$ also yields perfect correlations, i.e.\  $H(Z_A|Z_B)=0$. For the complementarity factor we have $q_{\rm MU}=\log(d)$ since $\abs{\braket{\mathbb{Z}^z}{\mathbb{X}^x}}=1/\sqrt{d}$ for all pairs of basis states, a property called mutual unbiasedness. Thus, Eq.~\eqref{eq:urelMU} yields $-H(A|B)\geq \log(d)$, which is a tight lower bound.

The example above is remarkable as measuring in only two different bases allows detecting all the entanglement that the state contains. However, measuring in MUBs is in general extremely challenging especially when considering subsystems containing multiple particles. The set of experimentally feasible local measurement pairs will be strongly restricted and depend on the system under study and thus Eq.~\eqref{eq:urelMU} will in general not be tight. Also, as we show in \cite{long_paper}, even assuming measurements in MUBs, the relation is only tight for very special states, such as the maximally entangled state, which may be far from the states that are of interest from the perspective of many-body physics. 
%In fact, in \cite{long_paper} we show that, if measuring in two bases connected by a Fourier transform, the maximally entangled state is the only pure state (besides separable states) for which Eq.~\eqref{eq:urelMU} is tight. 
The main goal of this work is to derive an improved uncertainty relation, that allows one to detect a significant amount of entanglement for experimentally relevant classes of states and under realistic assumptions about the available measurement bases.
One reason for the lack of tightness of Eq.~\eqref{eq:urelMU} is that the complementarity factor is state independent. It involves a maximization over all basis state overlaps and thus only depends on the chosen measurements.
For non-MUB measurements some overlaps may be large leading to a small $q_{\rm MU}$ and thus little detectable entanglement.
The key idea is to exploit that not all basis states contribute equally to a given state $\hat{\rho}_{AB}$ and thus one can tighten the bound by including measured state occupancies.

We are now ready to state the main result of this work: 
\begin{theorem}[State dependent entanglement bound] \label{fully_state_dep_bound}
	Let $\HS = \HS_A \otimes \HS_B$ be a bipartite Hilbert space. Let $X$ and $Z$ be two measurements in the bases $\{\xeket_A\}$ and $\{\zeket_A\}$ on $\HS_A$, and $X'$ and $Z'$ be measurements in bases on $\HS_B$. Let $\coverlap{x}{z} = \abs{\esbraket{z}{x}}^2$ and
	\begin{equation}
	    q_{\rm FSD} = - \sum_{x, x'} P_{XX'}(x,x') \log(\sum_z \coverlap{x}{z} P_{ZX'}(z|x')) \, ,
	\end{equation}
	where $ P_{XX'}(x,x')$ is the probability for measuring outcome $x$ in $A$ and $x'$ in $B$ upon measuring $X$ and $X'$, respectively, and $P_{ZX'}(z|x')$ is the conditional probability for obtaining outcome $z$ upon measuring $Z$ in $A$ given that outcome $x'$ was found upon measuring $X'$ in $B$.
    Then
	\begin{equation}\label{eq:urelFSD}
	    -H(A|B) \geq - H(X_A|X'_B) - H(Z_A |Z'_B) + q_{\rm FSD} \,.
	\end{equation}
	\begin{proof}
	A proof is given in Ref.~\cite{long_paper}.
	\end{proof}
\end{theorem}
The complementarity factor $q_{\rm FSD}$ is state dependent and no maximization over the basis state overlaps appears any more. Note that $q_{\rm FSD} \geq q_{\rm MU}$, and so for not mutually unbiased measurements, this improves the tightness compared to Eq.~\eqref{eq:urelMU} as well as to previously known state-dependent bounds \cite{coles_improved_2014}. The conditional probabilities $P_{ZX'}(z|x')$ appearing in $q_{\rm FSD}$ need to be determined through measurements. Thus, in addition to $P_{XX'}(x, x')$ and $P_{ZZ'}(z, z')$ a joint measurement of $Z$ on $A$ and $X'$ on $B$ is required. In the examples below we will always consider equal subsystems $ \HS_A = \HS_B$ and measurements $X=X'$ and $Z=Z'$.

\begin{figure}[t]
	\includegraphics[width=\columnwidth]{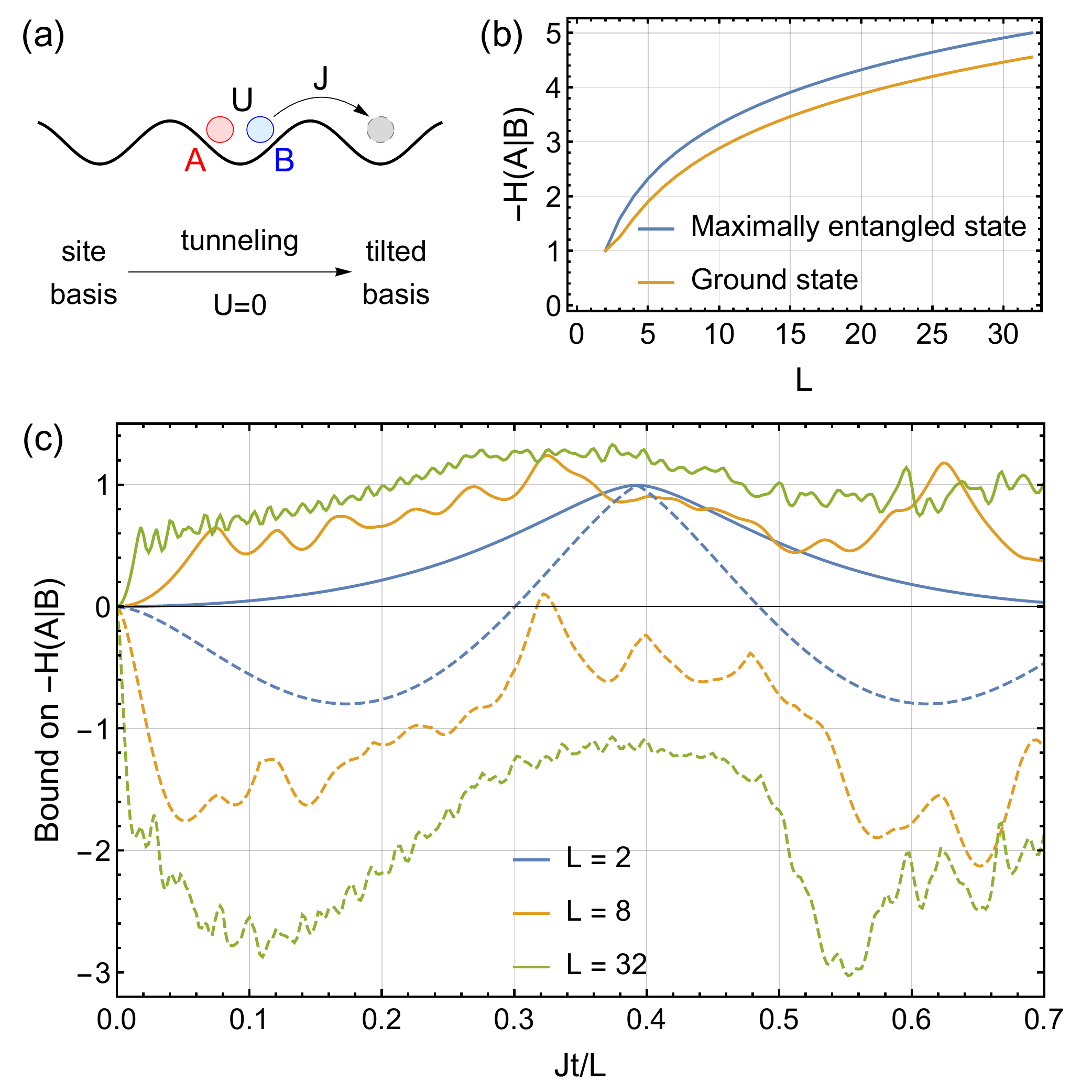}
	\caption{\textbf{Two particles on a lattice.} (a) Illustration of the system Hamiltonian and measurement bases. For the site basis measurement the positions of both particles on the lattice are detected. For the tilted basis measurement the particles are allowed to tunnel independently for time $t$ before their positions are measured. (b) Entanglement entropy of the ground state of two particles with strongly attractive interactions $U/J=-100$ compared to the maximally entangled state. (c) Detectable entanglement using Eq.~\eqref{eq:urelFSD} (solid lines) and Eq.~\eqref{eq:urelMU} (dashed lines) as a function of the tunneling time $t$.}
	\label{fig:lattice}
\end{figure}

\textit{Particles on a lattice.}--
Our first example system is a pair of distinguishable particles on an $L$-site lattice. The particles can tunnel between the lattice sites with equal rate $J$ and interact with interaction strength $U$ if they are on the same site of the lattice [cf.\ Fig.~\ref{fig:lattice}(a)]. Thus their dynamics is governed by the Hubbard Hamiltonian
\begin{equation}
    \hat{H}= -J\!\!\!\! \sum_{p\in \{A,B\}}\sum_{i = 1}^{L-1} (\hat{a}^\dagger_{p,i\/}\,\hat{a}_{p,i+1} + \text{h.c.} ) + U \sum_{i=1}^{L} \hat{n}_{A,i\/} \, \hat{n}_{B,i}
\end{equation}
where $\hat{a}^\dagger_{p,i}$ creates a particle of species $p$ on lattice site $i$ and $\hat{n}_{p,i\/}=\hat{a}^\dagger_{p,i}\, \hat{a}_{p,i}$. We consider entanglement between the two particles $A$ and $B$. The local Hilbert spaces are spanned by the states $\ket{i_p}$ of particle $p$ occupying site $i$. In the "site basis" $\ket{i_Ai_B}$, for strongly attractive interactions ($U<0$, $|U|/J\gg 1$), the ground state of this model is close to the maximally entangled state discussed above, i.e. a superposition of all states with both particles occupying the same lattice site. Figure~\ref{fig:lattice}(b) shows that the coherent information of the ground state for $U/J=-100$ increases as $\log(L)$. Compared to the maximally entangled state, the ground state features slightly less entanglement due to the fact that the sites close to the boundary of the lattice are energetically less favorable and thus have reduced occupations. The experimental preparation of such entangled ground states has been demonstrated for two sites \cite{bergschneider_experimental_2019}.

We show that entanglement can be detected using our improved entropic bound \eqref{eq:urelFSD} by choosing the site basis for the measurements $X$ and $X'$, experimentally realizable using standard tools \cite{Greiner2015, Boll2016, Schaefer2020}. A second "tilted basis" measurement is implemented by letting the particles tunnel independently for a time $t$ by switching off the interactions between them before measuring the particles' positions. We choose $Z$ and $Z'$ both to be measurements in the tilted basis.
As shown in Fig.~\ref{fig:lattice}(c) we only obtain a tight entanglement bound in the case of $L=2$ sites where at $tJ=\pi/4$ an MUB measurement is realized. For larger lattice sizes the entanglement bounds are far from tight, however, our improved relation does detect a significant amount of entanglement (solid lines) while previously known bounds fail to detect any entanglement (dashed lines).

\begin{figure}
	\includegraphics[width=\columnwidth]{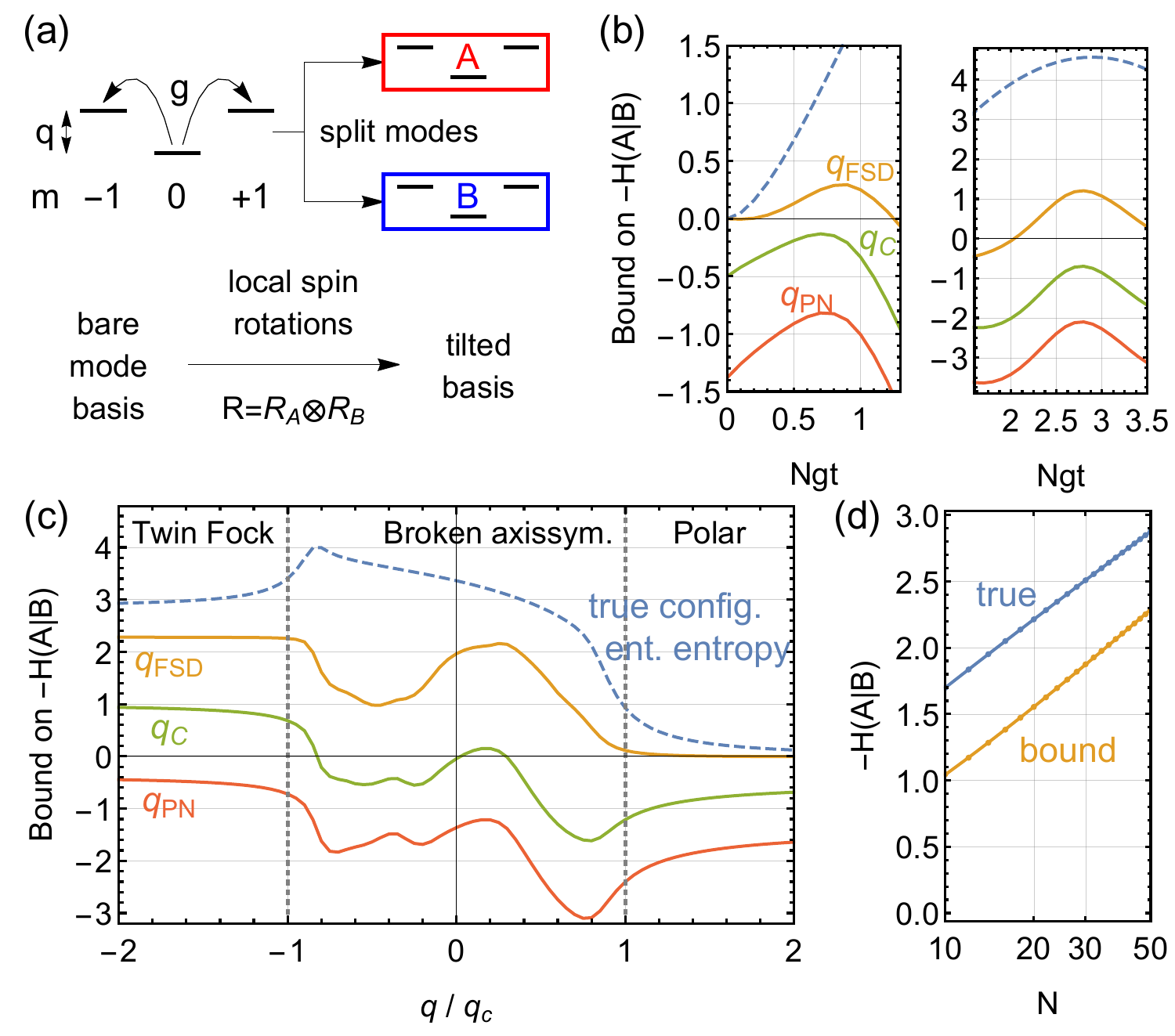}
	\caption{\textbf{Collective spin-1 system.} (a) Schematic of the system Hamiltonian and local measurement bases. Subsystems $A$ and $B$ are obtained by applying a balanced beam splitter operation to each mode. For the bare mode basis one detects the number of particles in each mode, while for the tilted basis occupation numbers are detected after a non-interacting local unitary was applied. (b) Entanglement generated by the evolution under spin-mixing dynamics out of the polar state $\ket{0,N,0}$ ($N=50$, $q=- g\,(N - 1/2)$) (dashed) and state dependent bounds. For short and long-time evolution different tilted bases were used (see text). (c) Entanglement and bounds for the ground state of Hamiltonian \eqref{eq:Hspin1} for varying $q$ ($N=50$). The dotted vertical lines designate the boundaries between different ground state phases. (d) Scaling of true entanglement entropy and bound using $q_{\rm FSD}$ deep in the Twin-Fock phase ($q/q_c=-5$). Only even $N$ are shown. For odd $N$, both $-H(A|B)$ and the bound take somewhat larger values but scale in the same way with $N$ \cite{long_paper}. The solid lines are fits to the data points. }
	\label{fig:spin1}
\end{figure}

\textit{Collective spin-1 system.}--
Our second example system is an ensemble of $N$ collectively interacting three-level systems, or, equivalently, bosonic particles in the ground state of some trapping potential. This choice is motivated by recent experiments demonstrating entanglement generated in spinor Bose-Einstein condensates (BECs) \cite{kunkel_spatially_2018, fadel_spatial_2018, lange_entanglement_2018}. These works used variance based inseparability criteria to detect the presence of entanglement. Here we show that atom number resolved measurements would enable the detection of bounds on entanglement measures.

For concreteness, we consider $N$ $^{87}$Rb atoms initially prepared in the internal state $m_F=0$ of the $F=1$ hyperfine manifold. This initial Fock state \footnote{Note that in realistic BEC experiments the total atom number $N$ fluctuates. This will in general affect our entanglement detection protocol, but if atom-number resolved measurements are available, post-selection on $N$ can be applied.} $\ket{N_{-1},N_0,N_1}=\ket{0,N,0}$ then evolves under the Hamiltonian \cite{Hamley2012, kunkel_spatially_2018, lange_entanglement_2018} 
\begin{equation}\label{eq:Hspin1}
 \hat{H} = g \hat{a}^\dagger_1 \hat{a}^\dagger_{-1} \hat{a}_0 \hat{a}_0 + {\rm h.c.} + [g(\hat{N}_0 -1/2)+q](\hat{N}_1 +\hat{N}_{-1}) \,,
\end{equation}
where the first term describes spin-changing collisions creating pairs of particles in the $\pm 1$ modes, while the second term accounts for the quadratic Zeeman effect (and elastic collisions), see Fig.~\ref{fig:spin1}(a). When $q$ is tuned such that the second term approximately vanishes, the short-time evolution creates a state reminiscent of a two-mode squeezed vacuum state \cite{Hamley2012, Peise2015}.

The system is split into parts $A$ and $B$ by coupling each mode to an ancillary mode resulting in the beam splitter transformation $\hat{a}^\dagger_k \rightarrow \left( \hat{a}^{\dagger}_{A, k} + \hat{a}^{\dagger}_{B, k} \right)/\sqrt{2} $. Experimentally, this can be realized by spatially splitting the atomic cloud \cite{kunkel_spatially_2018, fadel_spatial_2018} or by coupling to additional internal levels \cite{kunkel_splitting_2019}.
This type of division into subsystems comes with a peculiar feature encountered when dividing systems of indistinguishable particles into local subsystems, namely fluctuating local particle numbers. As a consequence the entanglement entropy decomposes into two parts. For pure states, considered in the present example, one has $- \qent{A|B} = \ent(\{p(n)\}) + \sum_n p(n) \qent{\hat{ρ}_B^{(n)}}$, where where $p(n)$ is the probability distribution of particle number $n$ in subsystem $A$ (or $B$) \cite{Lukin2019}. The first term is the particle number entanglement which is due to the splitting process only and is non-zero even in the absence of interactions. The second term is the configurational (or spin) entanglement which is created through interactions. The operations that we admit for transformations between local bases conserve the local particle number, and thus we only detect the configurational entanglement.
In Ref.~\cite{long_paper} we generalize these notions, also for mixed states, by observing that the configurational contribution to the quantum conditional entropy is obtained by removing any coherences between different (generalized) particle number sectors in the state $\hat{\rho}_{AB}$. 

As local measurements we allow the detection of the particle number in each mode ("bare mode basis") and the corresponding tilted bases reached by locally applying collective SU(3) spin rotations $\hat{R} = \exp(i \sum_{j, k} C_{jk} \hat{a}^\dagger_j \hat{a}_k)$ prior to detecting the mode occupation numbers. This choice is motivated by the experimental capabilities of typical BEC experiments \cite{kunkel_spatially_2018, kunkel_splitting_2019}. We note that the measurement bases reachable in this way are far from mutually unbiased, as MUB measurements would require unitary transformations generated by terms of higher order in the mode operators, and it is unclear how such transformations could be implemented experimentally.

As a consequence of particle number fluctuations between subsystems, the split state includes a component where all particles end up on one side. Thus the complementarity factor $q_{\rm MU}$ in Eq.~\eqref{eq:urelMU} vanishes since in this case any pair of operators has the vacuum as a common eigenstate.
This means that any non-trivial entanglement bound is necessarily state dependent in this case. Besides our fully state dependent bound, we consider two less state dependent bounds, which can be derived from an entropic uncertainty relation due to Coles and Piani \cite{coles_improved_2014}. The complementarity factor $q_{\rm PN}$ takes into account the measured total particle number in each subsystem, and $q_{\rm C}$ exploits the full distribution of outcomes of the $X$ measurement in subsystem $A$, see Ref.~\cite{long_paper} for details. Both of these bounds are strictly weaker than our fully state dependent bound ($q_{\rm PN}\leq q_{\rm C}\leq q_{\rm FSD}$).

In our numerical experiments we fixed $X$ (and $X'$) as the bare mode basis and chose $Z$ (and $Z'$) by maximizing the detected entanglement over all possible SU(3) rotations (see Ref.~\cite{long_paper} for details). We found that in almost all cases, choosing $Z$ to be reached by the "single particle Fourier transform" $[\exp(C)]_{jk}= (i/\sqrt{3}) \exp(2 \pi i j k/3)$ ($j,k = 0, 1, 2$) is the optimal choice. The entanglement generated at short times, shown in the left panel of Fig.~\ref{fig:spin1}(b), is detected by imprinting the phases $(\phi_1, \phi_0, \phi_{-1}) = (0.095, -0.495 , 0.400)\pi$ onto the modes before applying the single particle Fourier transformation. At later times zero phase imprint is optimal. In both cases our improved entanglement bound is far from tight but presents a large improvement over previously known bounds which do not detect any entanglement.

Next, we consider the ground state of the Hamiltonian \eqref{eq:Hspin1}, which features quantum phase transitions at $|q|=q_c=2N|g|$ \cite{feldmann_interferometric_2018}. In the polar phase ($q > q_c$), the ground state is close to $\ket{0,N,0}$ with little configurational entanglement. When entering the broken-axissymmetry phase ($|q| < q_c$), where all modes are populated, entanglement increases and is detected by our bound, as shown in Fig.~\ref{fig:spin1}(c). Finally, in the Twin-Fock phase $q < -q_c$), the ground state is close to $\ket{N/2,0,N/2}$. Most of of the configurational entanglement of this state is detected by our entropic uncertainty bound \eqref{eq:urelFSD}. Our bound clearly outperforms the previously known state-dependent bounds.
Figure~\ref{fig:spin1}(d) shows the $N$-dependence of the true and detected configurational entanglement in the Twin-Fock phase. While the true value scales as $(1/2)\log(N)$, fitting the bound with $a\log(N+b)+c$ yields $a=0.599(4)$. This means that the bound increases faster than the true value, leading us to the conjecture that the bound becomes tight asymptotically at large $N$.

\textit{Conclusions.}--
In conclusion, we have established an experimentally accessible lower bound on distillable entanglement and demonstrated its applicability to relevant experimental setups. We stress that our entanglement bound only requires measurements in two bases which do not need to be mutually unbiased. Thus we expect application across all experimental platforms for quantum simulation. In the future, it will be important to test the requirements in terms of measurement statistics for obtaining faithful estimates for conditional entropies and find ways to calculate complementarity factors analytically. Both these aspects will determine the degree to which this approach is scalable and will most likely depend on the considered system and states.

\acknowledgements
\textit{Acknowledgements.}-- We thank Stefan Flörchinger, Tobias Haas, Philipp Kunkel, and Markus Oberthaler for helpful discussions. This work is supported by the Deutsche Forschungsgemeinschaft (DFG, German Research Foundation) under Germany’s Excellence Strategy EXC2181/1-390900948 (the Heidelberg STRUCTURES Excellence Cluster) and within the Collaborative Research Center SFB1225 (ISOQUANT).

\nocite{guhne_entropic_2004}
\bibliography{refs}

\end{document}